\documentclass[sigconf]{acmart}

\usepackage[utf8]{inputenc}
\usepackage[skip=5pt]{caption}

\usepackage{graphicx}
\usepackage{amsmath}
\usepackage{booktabs}
\usepackage{CJKutf8}
\def\BibTeX{{\rm B\kern-.05em{\sc i\kern-.025em b}\kern-.08emT\kern-.1667em\lower.7ex\hbox{E}\kern-.125emX}}
    
\copyrightyear{2019} 
\acmYear{2019} 
\setcopyright{acmcopyright}
\acmConference[SIGIR '19]{Proceedings of the 42nd International ACM SIGIR Conference on Research and Development in Information Retrieval}{July 21--25, 2019}{Paris, France}
\acmBooktitle{Proceedings of the 42nd International ACM SIGIR Conference on Research and Development in Information Retrieval (SIGIR '19), July 21--25, 2019, Paris, France}
\acmPrice{15.00}
\acmDOI{10.1145/3331184.3331434}
\acmISBN{978-1-4503-6172-9/19/07}
\begin{document}

%
% The "title" command has an optional parameter, allowing the author to define a "short title" to be used in page headers.
\title[Applying Deep Learning in E-commerce Search]{From Semantic Retrieval to Pairwise Ranking: Applying Deep Learning in E-commerce Search}

\fancyhead{}
%
% The "author" command and its associated commands are used to define the authors and their affiliations.
% Of note is the shared affiliation of the first two authors, and the "authornote" and "authornotemark" commands
% used to denote shared contribution to the research.
\author{Rui Li, Yunjiang Jiang, Wenyun Yang, Guoyu Tang, Songlin Wang, Chaoyi Ma, Wei He, Xi Xiong, Yun Xiao, Eric Yihong Zhao}
\affiliation{%
   \institution{JD.com, Mountain View, CA}
%  \streetaddress{Anonymous address}
  % \city{}
  % \state{CA} \\
   \{rui.li, yunjiang.jiang, wenyun.yang, tangguoyu, wangsonglin3, machaoyi, hewei92, xiongxi, xiaoyun1, ericzhao\}@jd.com
}

%
% By default, the full list of authors will be used in the page headers. Often, this list is too long, and will overlap
% other information printed in the page headers. This command allows the author to define a more concise list
% of authors' names for this purpose.
\renewcommand{\shortauthors}{Rui Li et.al}
%
% The abstract is a short summary of the work to be presented in the article.
\begin{abstract}
We introduce deep learning models to the two most important stages in product search at JD.com, one of the largest e-commerce platforms in the world. Specifically, we outline the design of a deep learning system that retrieves semantically relevant items to a query within milliseconds, and a pairwise deep re-ranking system, which learns subtle user preferences. Compared to traditional search systems, the proposed approaches are better at semantic retrieval and personalized ranking, achieving significant improvements.
\end{abstract}

\settopmatter{printacmref=false}

\ccsdesc[500]{Information systems~Information retrieval}
%\ccsdesc[500]{Computing methodologies~Neural networks}

%
% Keywords. The author(s) should pick words that accurately describe the work being
% presented. Separate the keywords with commas.
\keywords{Neural networks; Semantic Search; Personalized Ranking}

%
% A "teaser" image appears between the author and affiliation information and the body 
% of the document, and typically spans the page. 

%
% This command processes the author and affiliation and title information and builds
% the first part of the formatted document.
\maketitle

\newcommand{\header}[1]{{\flushleft \textbf{#1}}}
\newcommand{\sheader}[1]{{\flushleft \textit{#1}}}

\newcommand{\eg}{\textit{e.g.}}
\newcommand{\xeg}{\textit{E.g.}}
\newcommand{\ie}{\textit{i.e.}}
\newcommand{\etc}{\textit{etc.}}
\newcommand{\etal}{\textit{et al.}}
\newcommand{\wrt}{\textit{w.r.t.}} 

\section{Introduction}
Over past decades, online shopping platforms have become ubiquitous in people's daily life. In a typical e-commerce search engine, two major stages are involved to answer a user query, namely \textbf{Candidate Retrieval}, which uses inverted indexes to efficiently retrieve candidates based on term matching, and \textbf{Candidate Ranking}, which orders those candidates based on factors, such as relevance and predicted conversion ratio. Here we share our experiences and promising results of applying deep learning for both stages in e-commerce search at JD.com. 

For candidate retrieval, we introduce the Deep Semantic Retrieval (DSR) system, in addition to traditional term matching based retrieval systems. DSR encodes queries and items into a semantic space based on click logs and human supervision data, and uses an efficient K-nearest neighbor search algorithm to retrieve relevant items from billions of candidates within milliseconds. For candidate ranking, we introduce the Deep Pairwise Ranking (DPR) system to leverage both hand-crafted numerical features and raw sparse features. DPR uses a Siamese architecture to effectively learn preference between a pair of items under the same query from billions of search log data and efficiently score each item individually (without enumerating all the pairs) online. 

\section{Deep Semantic Retrieval}
\label{sec:overview}
The most critical challenge for this stage is enabling semantic retrieval beyond exact term matches. Based on our analysis, this problem impacts around $20\%$ of our search traffic. 

\begin{figure}
    \centering
    \includegraphics[width=75mm]{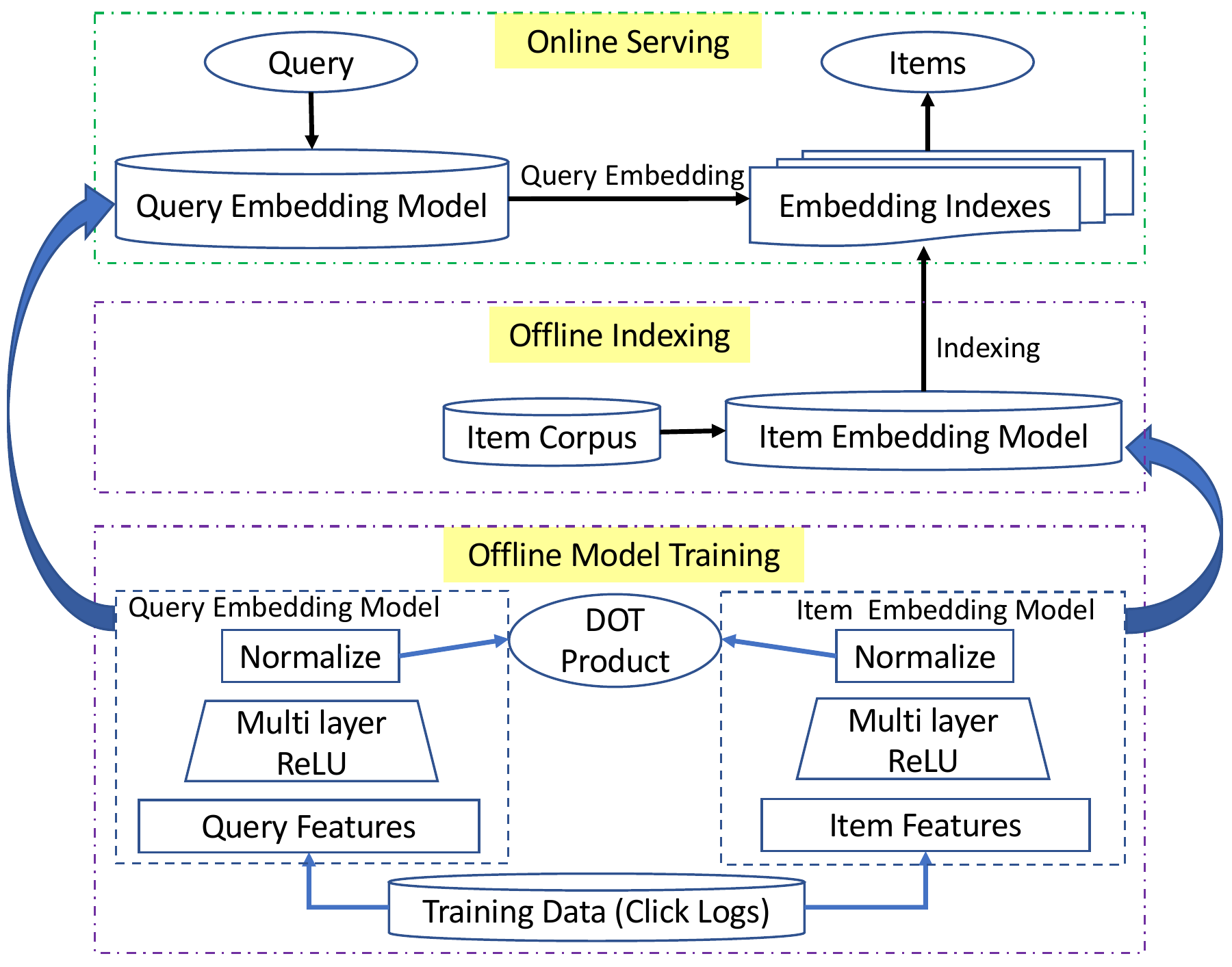}
    \vspace{-2mm}
    \caption{Overview of deep semantic retrieval system.}
    \label{fig:overview}
\vspace{-5mm}
\end{figure}

\header{System Overview}  Neural network models naturally address this problem via semantic term embeddings. While many deep neural network models (\eg, DSSM~\cite{Huang:2013:LDS:2541176.2505665}) have been proposed for search relevance ranking, our DSR system explores the rare territory of candidate retrieval. As Figure~\ref{fig:overview} shows, DSR has three parts.
\vspace{-1mm}
\sheader{Offline Training} process learns a query, item two tower model. Each tower consists of 1) an input layer that concatenates input features of a query or an item, 2) an encoding ReLU multi-layer, and 3) a normalization layer to ensure L2 normalized output embeddings.
\vspace{-1mm}
\sheader{Offline Indexing} module computes all the item embeddings from the item corpus, then builds an embedding index offline for efficient online embedding retrieval.

\sheader{Online Serving} module loads the query part of the two tower model and computes an embedding for any query, which then retrieves the $K$ nearest/most similar items from the item embedding index. 

\label{sec:model}
\header{Model Overview} Now, we give a overview of our proposed model. Given a training set $\mathcal{D} = \{(q_i, s^+_j, s^-_k)\;|\; i, j, k\}$,
where $q_i$ stands for a query, $s^+_j$ stands for a positive item that is relevant to $q_i$, and $s^{-}_k$ stands for a negative item that is irrelevant to $q_i$, we learn the query and item towers jointly by minimizing the following loss function
\[
L(\mathcal{D}) = \sum_{(q_i, s_j^+, s_k^-) \in \mathcal{D}} \max\left(0, \delta - \left(Q(q_i)^T S(s_j^+) - Q(q_i)^T S(s_k^-)\right)\right)
\]
where $\delta \geq 0$ is the margin used in hinge loss, $Q(\cdot)$ denotes the query tower, $S(\cdot)$ denotes the item tower and the superscript $\cdot^T$ denotes transpose operation. 
Intuitively, this function employs the hinge loss to encourage the dot product between a query $q_i$ and a positive item $s_j^+$ to be larger than the product between the query $q_i$ and a negative item $s_k^-$ by a certain margin $\delta$. 

\header{Experimental Results}
Table~\ref{tab:retrieval_abtest} shows relative improvement of DSR in online A/B tests in terms of four core business metrics, including user click through rate (UCTR), user conversion rate (UCVR), and gross merchandise value (GMV), as well as query rewrite rate (QRR), which is believed to be a good indicator of search satisfaction.  Table~\ref{tab:gpu} reports the time consumed by DSR for indexing and searching $120$ million items with Nvidia Tesla P40 GPU and Intel $64$-core CPU: GPU achieves \textasciitilde10x speed up. Table~\ref{tab:good_cases} shows a few sample queries to better illustrate the power of DSR at bridging semantic gaps between queries and relevant items.

\begin{table}[tb]
    \centering
    \begin{tabular}{c|c c c c}
    \hline
            & UCTR  & UCVR  & GMV  & QRR \\
    \hline
    overall & $+1.61\%$ & $+0.98\%$ & $+0.54\%$ & $-4.34\%$   \\
    long tail & $+9.7\%$  & $+10.03\%$ & $+7.5\%$ & $-9.99\%$ \\
    \hline 
    \end{tabular}
    \caption{Online A/B test improvements.}
    \label{tab:retrieval_abtest}
\vspace{-5mm}
\end{table}

\begin{table}[tb]
    \centering
    \begin{tabular}{c|c c c}
    \hline
        &  indexing (sec.)  &  search (ms) & QPS\\
    \hline
    CPU   & $3453$   & $9.92$ & $100$\\
    GPU  & $499$  & $0.74$ & $1422$ \\   
    \hline
    \end{tabular}
    
    \caption{Efficiency for indexing and serving.}
    \label{tab:gpu}
\vspace{-4mm}
\end{table}

\begin{CJK*}{UTF8}{gbsn}
\begin{table}[bt]\small{
    \centering
    \begin{tabular}{c|cc}
    \hline
    query & retrieved item \\
    \hline
     奶粉\;大童    &  美赞臣\;安儿健A+\;4段 \\
     (milk powder big kid) & $\left(\begin{minipage}{1.5in}Enfamil A+ level-4 for 3-6 yr old \end{minipage}\right)$ \\
% \hline
 %    送爷爷的手机    &  老人手机$\;$三防手机\\
 %    \left(\begin{minipage}{1in}gift cellphone for grandpa\end{minipage}\right) & %\left(\begin{minipage}{1.5in}senior cellphone, water dust shake proof %cellphone\end{minipage}\right) \\
     \hline
     学习自由泳器材    & 英发/yingfa 划臂\\
     $\left(\begin{minipage}{1in}
     learn free-style swimming equipment
     \end{minipage}\right)$ & (yingfa hand paddle ) \\
    \hline
    \end{tabular}
    \caption{Good cases from DSR}
    \label{tab:good_cases}
\vspace{-7mm}}
\end{table}
\end{CJK*}

\section{Deep Pairwise Ranking}
Traditional e-commerce search systems apply gradient boosted trees (xgboost) models~\cite{chen2016xgboost} to rank products based on predicted conversion rates. Relying mainly on numeric features, those models cannot effectively explore sparse features, such as user's click/purchase history. Deep learning(DL) naturally address this problem as it can leverage those raw features via embeddings. However, to apply DL in ranking, we need to address 1) big computation overhead and 2) sufficient amount of good training data required by DL.  %Due to the computational overhead, we only score the top K (100) items selected from earlier phases with a complex DL model, as we observed that 90\% of clicks/orders are within top 75 positions in existing search results.

\header{Overview} To address those challenges, we first introduce a \emph{re-ranking} stage, which only scores the top K (100) items selected from earlier phases with a complex DL model, to reduce the required computation, as we observe that 90\% of clicks/orders are within top 75 positions in existing results. Second, we choose to train a pairwise ranking model with a large amount of noisy logs. While the top 100 results are usually all reasonably relevant, the focus is personalized preference, which is hard to quantify. Logs, though abundant, are only good for approximating theses subtle preferences among presented items. Thus, instead of learning conversion rate, as most e-commerce ranking models do, we learn user preference between a pair of items for a given query.

\begin{figure}
    \includegraphics[width=0.46\textwidth]{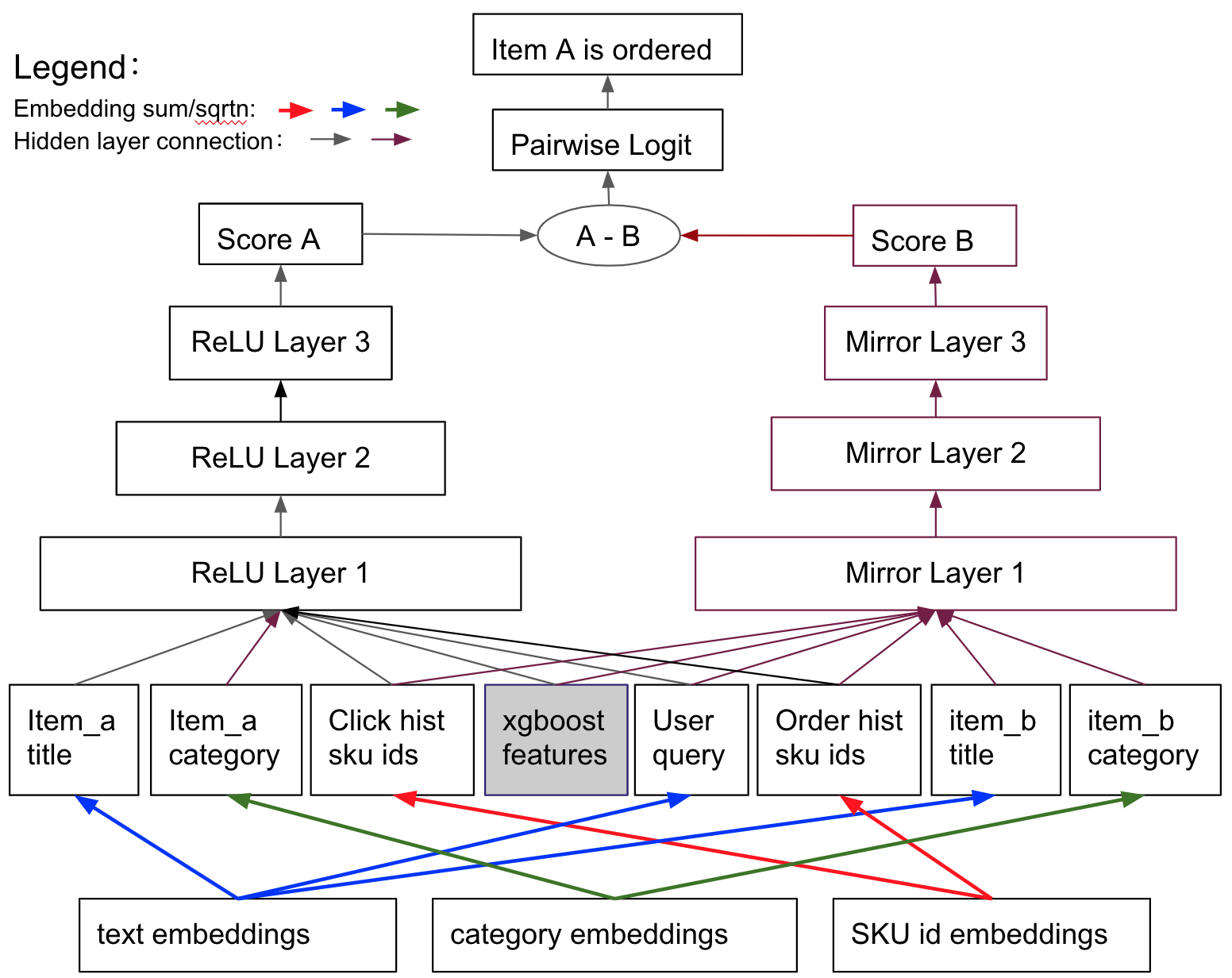}
    \caption{Simplified Siamese ranking model.}
    \label{fig:reranking}
\vspace{-3mm}
\end{figure}
\header{Model Details} In order to effectively learn preference between a pair of items under the same query offline and efficiently score each item individually (without enumerating all the pairs) online, we design a pairwise Siamese model, as shown in Figure~\ref{fig:reranking}. The training data consists of triples $(q, a, b)$ where $a, b$ are two items co-occurring under the same session with query $q$. The Siamese structure has two identical towers (sharing the same parameters) for item $a$ and item $b$.  Each tower takes all the input features from user, query, and item, goes through 3 ReLU layers, and outputs a logit.  The difference of the two logits is then combined with the binary label in a cross-entropy loss function. During online serving, only one of the two identical towers is used to score each item.

The model uses the following features: 1) hand picked numeric features, such as user's purchasing power and past CTR/CVR/sale volume, 2) raw text features for queries and items, tokenized into unigrams and bigrams, and 3) raw user action features, such as historical click streams. The latter two use sum-pooled embeddings. 

\header{Experimental Results}
Our top-line validation metric is order-based average session AUC. Our DPR model, which uses not only all the Xgboost features (about 130 total) but also sparse features, performs significantly better than Xgboost methods (Table~\ref{tab:validation}). 
Table~\ref{tab:abtest} shows results of online A/B tests. DPR significantly improves all the core business metrics, including UCTR, UCVR, GMV, and average order position (AOP). As DPR uses orders as positive examples during our training, it achieves the most gain on UCVR. However, similar gains on UCTR/GMV are also observed.

\begin{table}
    \centering
    \begin{tabular}{c| c c}
    \hline
        & session AUC   & order NDCG@5 \\
    \hline
    Xgboost & 0.852    & 0.522 \\
    Siamese DL & 0.870   & 0.573 \\
    \hline
    \end{tabular}
    \caption{Offline validation metrics.}
    \label{tab:validation}
\vspace{-4mm}
\end{table}

 \begin{table}
    \centering
    \begin{tabular}{c|c c c c}
    \hline
            & UCTR  & UCVR  & GMV  & AOP \\
    \hline
    overall & $+1.16\%$ & $+1.99\%$ & $+1.04\%$ & $-2.10\%$   \\
    \hline 
    \end{tabular}
    \caption{Online A/B test improvements.}
    \label{tab:abtest}
\vspace{-6mm}
\end{table}

\bibliographystyle{ACM-Reference-Format}
\bibliography{bibliography}


\begin{thebibliography}{2}

%%% the Web address.

\ifx \showCODEN    \undefined \def \showCODEN     #1{\unskip}     \fi
\ifx \showDOI      \undefined \def \showDOI       #1{#1}\fi
\ifx \showISBNx    \undefined \def \showISBNx     #1{\unskip}     \fi
\ifx \showISBNxiii \undefined \def \showISBNxiii  #1{\unskip}     \fi
\ifx \showISSN     \undefined \def \showISSN      #1{\unskip}     \fi
\ifx \showLCCN     \undefined \def \showLCCN      #1{\unskip}     \fi
\ifx \shownote     \undefined \def \shownote      #1{#1}          \fi
\ifx \showarticletitle \undefined \def \showarticletitle #1{#1}   \fi
\ifx \showURL      \undefined \def \showURL       {\relax}        \fi
% invisible to TeX
\providecommand\bibfield[2]{#2}
\providecommand\bibinfo[2]{#2}
\providecommand\natexlab[1]{#1}
\providecommand\showeprint[2][]{arXiv:#2}

\bibitem[\protect\citeauthoryear{Chen and Guestrin}{Chen and Guestrin}{2016}]%
        {chen2016xgboost}
\bibfield{author}{\bibinfo{person}{Tianqi Chen} {and} \bibinfo{person}{Carlos
  Guestrin}.} \bibinfo{year}{2016}\natexlab{}.
\newblock \showarticletitle{Xgboost: A scalable tree boosting system}. In
  \bibinfo{booktitle}{\emph{KDD}}. ACM, \bibinfo{pages}{785--794}.
\newblock


\bibitem[\protect\citeauthoryear{Huang, He, Gao, Deng, Acero, and Heck}{Huang
  et~al\mbox{.}}{2013}]%
        {Huang:2013:LDS:2541176.2505665}
\bibfield{author}{\bibinfo{person}{Po-Sen Huang}, \bibinfo{person}{Xiaodong
  He}, \bibinfo{person}{Jianfeng Gao}, \bibinfo{person}{Li Deng},
  \bibinfo{person}{Alex Acero}, {and} \bibinfo{person}{Larry Heck}.}
  \bibinfo{year}{2013}\natexlab{}.
\newblock \showarticletitle{Learning deep structured semantic models for web
  search using clickthrough data}. In \bibinfo{booktitle}{\emph{CIKM}}.
  \bibinfo{pages}{2333–2338}.
\newblock


\end{thebibliography}

% 
% If your work has an appendix, this is the place to put it.
\end{document}